\documentclass[conference]{IEEEtran}
\usepackage[letterpaper, top=1.91cm, bottom=2.45cm, left=1.62cm, right=1.62cm]{geometry}
\usepackage{amsmath,amsfonts}
\usepackage{array}
\usepackage[caption=false,font=footnotesize,labelfont=rm,textfont=rm]{subfig}
\usepackage{textcomp}
\usepackage{stfloats}
\usepackage{url}
\usepackage{verbatim}
\usepackage{graphicx}
\usepackage{cite}
\usepackage{amsthm}
\usepackage{amssymb}
\usepackage{upgreek}
\usepackage{bm}
\usepackage[utf8]{inputenc}
\usepackage{diagbox}
\usepackage{xcolor}
\usepackage{multirow}
\usepackage{booktabs}
\usepackage{tabularx}
\usepackage{cleveref}
\usepackage{mathtools}
\usepackage{color}

\usepackage[ruled,vlined,linesnumbered]{algorithm2e}
\SetKwInput{KwInit}{Init}
\SetKwInOut{KwRequire}{Require}
\SetKwInOut{KwEnsure}{Ensure}
\SetKwComment{Comment}{// }{}

\SetCommentSty{mycommfont}

\begin{document}

\title{DiP-SD: Distributed Pipelined Speculative Decoding for Efficient LLM Inference at the Edge}

\author{
    \IEEEauthorblockN{Yaodan Xu, Sheng Zhou, Zhisheng Niu}
    \IEEEauthorblockA{Department of Electronic Engineering, Tsinghua University, Beijing 100084, P.R. China}
    \IEEEauthorblockA{xyd21@mails.tsinghua.edu.cn, \{sheng.zhou, niuzhs\}@tsinghua.edu.cn}
}

\maketitle

\begin{abstract}
Speculative decoding has emerged as a promising technique for large language model (LLM) inference by accelerating autoregressive decoding via draft-then-verify. This paper studies a new edge scenario with multi-user inference, where draft tokens are generated locally on devices and subsequently offloaded to a centralized edge server for batch verification. The key challenge is to sustain high throughput under coupled decisions of (i) batching and pipeline scheduling and (ii) per-user draft token length.
We propose DiP-SD, which exploits two complementary parallelism dimensions: device-level distributed drafting and phase-level draft--verify pipelining. We formulate a throughput-maximization objective, defined as the expected number of accepted tokens per unit time, and jointly optimize the number of batches, user-to-batch assignment, and integer draft lengths. To solve the resulting fractional mixed-integer program, DiP-SD scans the batch number and iteratively alternates between an association subproblem and a draft-length subproblem.
Numerical results under a Qwen3-1.7B/Qwen3-32B device--edge deployment show that DiP-SD achieves up to $17.89\times$ throughput over autoregressive decoding (AD) and $1.93\times$ over AD with greedy batching.
\end{abstract}


\section{Introduction}

Large language models (LLMs) are now widely used in interactive applications such as assistants, search, coding copilots, and customer support. These services repeatedly invoke \emph{LLM inference} to generate responses, making throughput and latency first-order system concerns. However, modern LLMs are parameter-heavy and resource-intensive, creating a fundamental tension between high inference cost and low-latency requirements for interactive use.

LLM inference is usually decomposed into a prefill stage and an autoregressive decode stage. While prefill can be amortized by caching and batching, decode remains latency-critical because tokens must be generated sequentially, making total latency proportional to output length. This token dependency makes decode the main throughput bottleneck for interactive services.
Speculative decoding (SD) addresses this bottleneck by using a draft-then-verify scheme: a fast draft model proposes multiple draft tokens, and a stronger target model verifies them in parallel. Foundational SD work has shown that carefully designed accept/reject-and-resample procedures can preserve the target distribution while generating multiple tokens per target model call~\cite{leviathan2023specdec,chen2023specsampling,xia2023specdec}.

Recent SD research improves acceleration from several directions. One line improves draft-target alignment and acceptance through distillation or online adaptation \cite{zhou2024distillspec,liu2024osd}. Another line redesigns the draft and verification structure, e.g., feature-space drafting, dynamic draft trees, or multi-head or self-speculative variants \cite{li2024eagle,li2024eagle2,cai2024medusa,zhang2024draftverify}. As SD moves to distributed deployment, communication and synchronization become dominant concerns, including adaptive window control, selective feedback, and bandwidth-aware sparsification \cite{yu2025dsd,ning2025dssd,zheng2025tkslt,bhattacharjee2025conformal}. 

Deploying inference at the edge can further improve responsiveness by placing models closer to users and reducing end-to-end delay. SD has also been explored in edge-oriented systems with heterogeneous devices and networks \cite{li2025sled,chen2025spin,venkatesha2025earlyexit,liu2025flowspec,li2026flexspec}, but a key caveat is that SD does not automatically scale to multi-user, high-concurrency workloads: throughput can degrade when verification becomes a shared bottleneck\cite{liu2024turbospec_goodput,chen2025spin}. This makes batching unavoidable, where batching means grouping multiple inputs to the same model to be processed together so that compute and overhead can be amortized to increase throughput; however, existing edge SD systems typically adopt simple batching policies, e.g., static batching in SLED \cite{li2025sled} or heuristic batch-size decisions in SPIN \cite{chen2025spin}. Beyond batching, pipeline parallelism has been studied to reduce bubbles and improve utilization: the authors of \cite{zhu2025efficient} pipeline draft and target phases via micro-batches within each draft--verify round, while FlowSpec \cite{liu2025flowspec} and PEARL \cite{liu2025pearl} pipeline draft and target execution within a single request.

\begin{figure}[t]
\centering
\includegraphics[width=0.90\columnwidth]{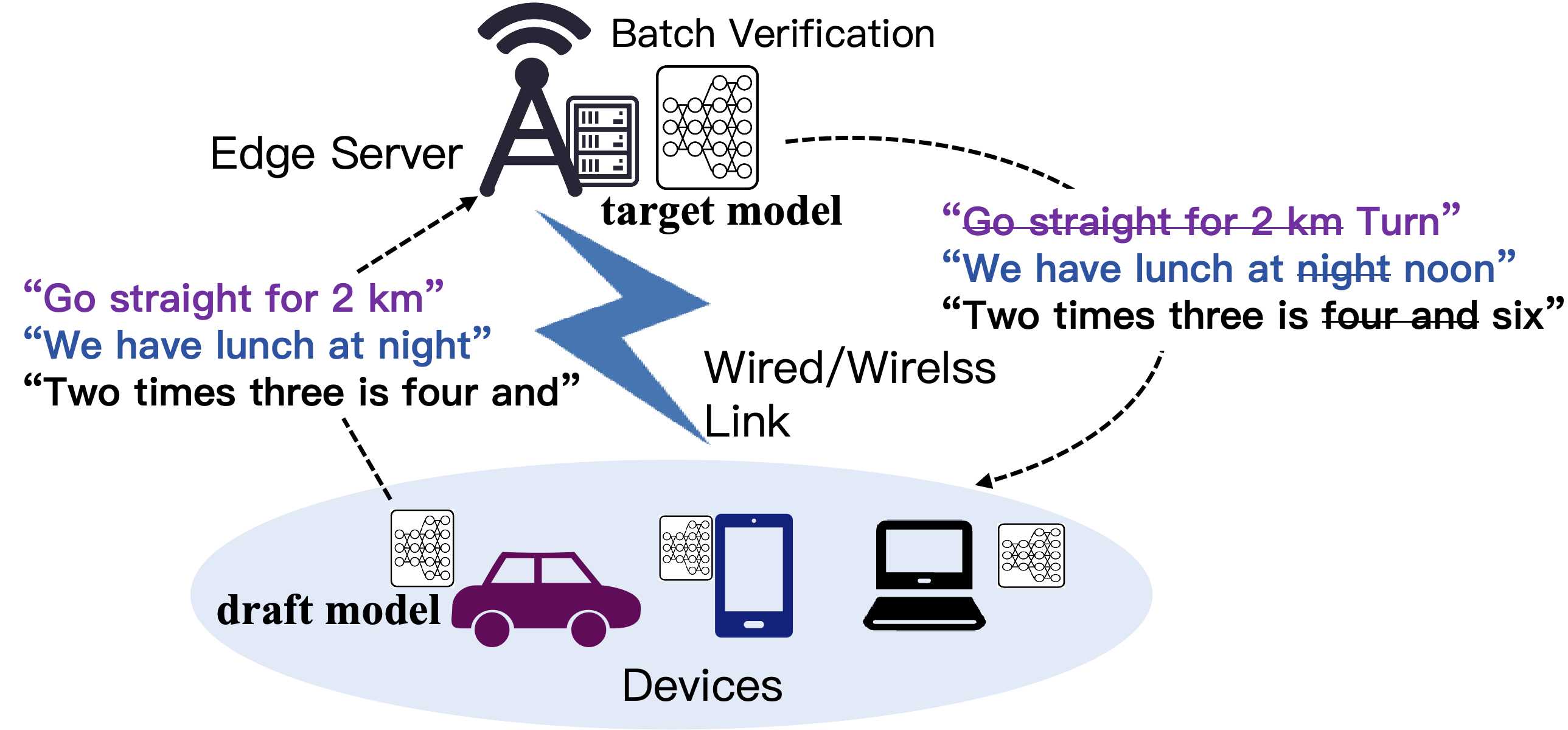}
\caption{Edge deployment of distributed pipelined speculative decoding: user-side devices run the draft model, and the edge server runs the target model for verification.}
\label{fig:wsd_setting}
\end{figure}


In this paper, we focus on a distributed edge setting where draft generation runs on multiple user-side devices locally and verification runs on an edge server, as shown in Fig.~\ref{fig:wsd_setting}. This setting is increasingly practical because small LLMs have been deployed on end devices (e.g., tablets and smartphones), enabling local drafting with server-side verification. It creates two coupled parallelism dimensions: device-level distributed draft parallelism and phase-level draft--target pipeline parallelism. Unlike \cite{zhu2025efficient}, which pipelines micro-batches within each draft--verify round, we pipeline across multiple batches over successive rounds, combining distributed drafting with batch-level verification pipelining.

DiP-SD is the first SD scheme to exploit batch-level pipeline parallelism between distributed drafting and centralized verification, with co-optimized speculation length and batching. 
The key control variables are the user-to-batch assignment, per-user speculative length, and the number of batches, which are tightly coupled by latency, memory, and communication constraints. Although DiP-SD is motivated by edge inference, it also applies to other SD deployments with distributed drafting and shared verification.

To address this coupled optimization problem, we propose a joint optimization framework. For each candidate number of batches, we iteratively alternate between a batch association subproblem and a draft-length subproblem, and select the best solution over scanned batch counts. Extensive numerical experiments show that the proposed scheme exhibits superior performance across all test points.

\section{System Model and Problem Formulation}

\subsection{Edge Deployment Scenario}


We consider an edge inference system serving $M$ user requests, indexed by $\mathcal{M}=\{1,\ldots,M\}$. The system adopts a distributed speculative decoding framework: lightweight draft models are deployed on distributed edge devices, while a large-scale target model (the verifier) resides on a centralized edge server.
As illustrated in Fig.~\ref{fig:wsd_setting}, the draft and verification phases are decoupled across heterogeneous compute resources. This architecture naturally enables device-level parallelism during drafting and phase-level pipelining between the devices and the server. Specifically, in each speculation round, each device sequentially generates a sequence of draft tokens, which are subsequently transmitted to the edge server for concurrent batch verification. The target model processes these tokens in a single forward pass, and employs techniques such as rejection sampling to determine their validity.

The iterative speculation process for each request follows a four-stage cycle:
\begin{enumerate}
\item \textbf{Local Drafting:} Each edge device independently employs its lightweight draft model to generate a sequence of tokens in a serial manner. These draft tokens are subsequently offloaded to the edge server.
\item \textbf{Batch Verification:} The edge server validates the received draft tokens through the target model in a single forward pass. It identifies the first mismatched token for each request, discards all subsequent incorrect predictions, and simultaneously produces one bonus token following the last accepted position, ensuring at least one token is generated per round. In our system, multiple requests are grouped into batches and verified together to amortize the overhead of target model inference.
\item \textbf{Synchronization:} The server transmits the verified tokens and the bonus token back to the respective device, ensuring both sides are synchronized with the latest generation state.
\item \textbf{State Update:} The device updates its local KV cache and resumes the next drafting round, starting from the newly received bonus token.
\end{enumerate}

\begin{figure}[t]
\centering
\includegraphics[width=0.98\columnwidth]{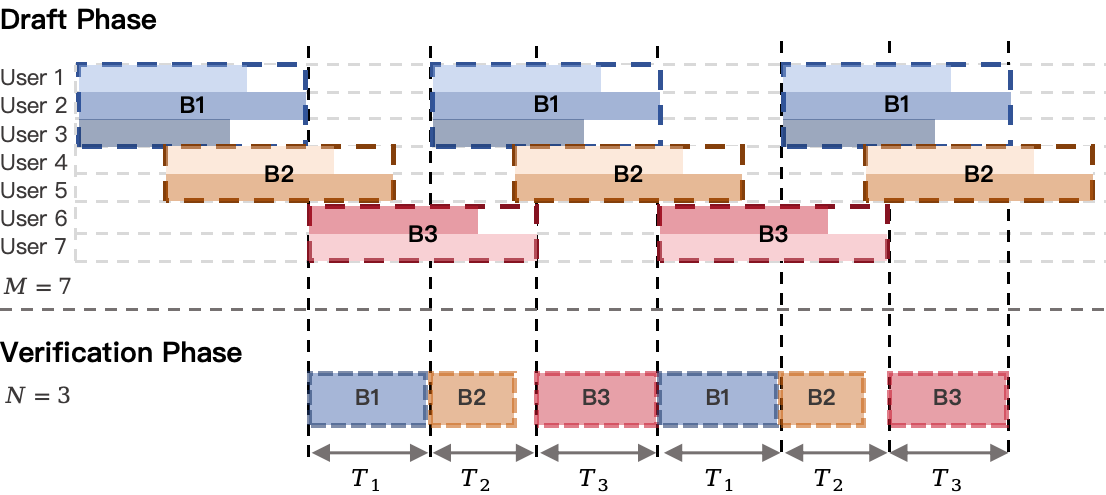}
\caption{Example of batching and pipelined verification. Requests are assigned into $N$ batches; each batch is verified at the edge server in sequence while drafting proceeds in parallel at devices.}
\label{fig:pipeline_batches}
\end{figure}

For the optimization, we introduce a batch count $N$ with index set $\mathcal{N}=\{1,\ldots,N\}$, where $N\in\{2,\ldots,M\}$. We use ``batch'' to denote a set of requests verified together at the edge server. Verification proceeds in $N$ pipeline \emph{stages}, where stage $n$ verifies batch $n$ (thus $n$ also indicates the verification order/slot). Fig.~\ref{fig:pipeline_batches} illustrates one example.

Request $m$ has prefix length $i_m$, communication overhead $\tau_m^{\mathrm c}$, and acceptance parameter $\alpha_m$. Here $\tau_m^{\mathrm c}$ is the per-round communication latency between the device and the edge server. The communication link can be wireless or wired, and the SD payload only contains drafted tokens and (when needed by the verification protocol) a small amount of probability information (e.g., draft token log-probabilities and sparse target logits), so the transmitted volume is small and we consider $\tau_m^{\mathrm c}$ as a constant.
The prefix length $i_m$ counts both the original input prompt and the previously accepted tokens (i.e., the current context length at the beginning of the optimization window).

We focus on optimizing the speculative decoding (decode) process and omit prefill latency, since prefill is executed once per request and can be treated as an additive constant. Since each request spans hundreds of tokens and thus many SD rounds, the set of active requests and system parameters are approximately stationary within each optimization horizon. Although the prefix length $i_m$ grows by the number of accepted tokens each round, this increment is small relative to the full generation length, so we treat $i_m$ as a constant.

\begin{table}[t]
\caption{Main notation in the system model.}
\label{tab:notation}
\centering
\begin{tabular}{ll}
\toprule
Symbol & Meaning \\
\midrule
$M,\mathcal{M}$ & user number and user index set \\
$N,\mathcal{N}$ & batch number and batch index set \\
$i_m$ & prefix length of user $m$ \\
$l_m$ & draft length of user $m$ \\
$\alpha_m$ & expected token acceptance rate of user $m$ \\
$x_{mn}$ & assignment variable from user $m$ to batch $n$ \\
$b_n$ & batch size (number of users in batch $n$) \\
$L_n,I_n$ & max draft length and max prefix length in batch $n$ \\
$t_n^{\mathrm d},t_n^{\mathrm v}$ & draft completion time and verification latency\\
$T_n$ & stage duration \\
$c_m,\beta_m,c^{\mathrm v},\beta^{\mathrm v}$ & linear latency coefficients \\
$J_{\Xi},h_1^{\Xi},h_2^{\Xi}$ & decoder blocks / hidden dims for model $\Xi\in\{\mathrm d,\mathrm v\}$ \\
\bottomrule
\end{tabular}
\end{table}

\subsection{Speculative Decoding Utility Model}
Each user $m$ selects integer draft length $l_m\in\mathbb{N}_{+}$.
The token acceptance rate $\alpha_m\in(0,1)$ characterizes the statistical alignment between the draft and target models. 
Given that this rate inherently fluctuates across different tasks, we model the acceptance rate as a user-specific parameter to account for such heterogeneity. 
The expected number of accepted tokens of one speculative round for user $m$ is
\begin{align}
u_m(l_m)=\frac{1-\alpha_m^{l_m+1}}{1-\alpha_m}, \label{eq:u_single}
\end{align}
where each draft token is accepted independently with probability $\alpha_m$
(geometric acceptance model~\cite{leviathan2023specdec}), and one bonus token
is always generated regardless of acceptance outcomes.
The total expected accepted tokens across all users is
\begin{align}
U(\mathbf{l})=\sum_{m=1}^{M}u_m(l_m). \label{eq:U_def}
\end{align}

\subsection{Transformer-Cost Primitives}
Let $\Xi\in\{\mathrm d,\mathrm v\}$ denote draft and verification (target) models. Model $\Xi$ has $J_{\Xi}$ decoder blocks, hidden dimension $h_1^{\Xi}$, and hidden dimension $h_2^{\Xi}$ in feed-forward network (FFN).

Following standard decoder-only cost scaling, we define
\begin{align}
F^{\mathrm d}(i)&=4J_{\mathrm d}h_1^{\mathrm d}\!\left(2h_1^{\mathrm d}+i+1+h_2^{\mathrm d}\right), \label{eq:Fd}\\
F^{\mathrm v}(L,I)&=4J_{\mathrm v}h_1^{\mathrm v}L\!\left(2h_1^{\mathrm v}+I+L+h_2^{\mathrm v}\right). \label{eq:Fv}
\end{align}
Intuitively, $F^{\mathrm d}(\cdot)$ and $F^{\mathrm v}(\cdot,\cdot)$ capture the compute intensity of draft and verification for one round, growing with the context length $i$ and the batch-wise maximum draft length and prefix length $(L,I)$.
Following~\cite{zhu2025efficient}, we model the per-step drafting and
verification latency as affine functions of the product $bF$:
\begin{align}
G_m(b,F)&=c_m bF+\beta_m, \label{eq:Gm}\\
G^{\mathrm v}(b,F)&=c^{\mathrm v} bF+\beta^{\mathrm v}. \label{eq:Gv}
\end{align}
The functions $G_m(\cdot,\cdot)$ and $G^{\mathrm v}(\cdot,\cdot)$ characterize the latency for drafting and verification, respectively. 
Parameters $(c,\beta)$ are obtained from profiling and capture per-compute scaling and fixed overhead, and $b$ is the batch size.
This affine dependence on batch size is consistent with empirical
observations reported in~\cite{10829781,xu2023smdp}.

\subsection{Assignment and Batch-Aggregation Constraints}
Binary assignment and one-batch-per-user constraints are
\begin{align}
x_{mn}\in\{0,1\},\quad \forall m,n, \label{eq:C_x_binary}\\
\sum_{n=1}^{N}x_{mn}=1,\quad \forall m. \label{eq:C_x_assign}
\end{align}
Here $x_{mn}$ indicates whether user $m$ is associated with batch $n$ (and thus verified in stage $n$).
Draft lengths are positive integers:
\begin{align}
l_m\in\mathbb{N}_{+},\quad \forall m. \label{eq:C_l_integer}
\end{align}

Batch size is
\begin{align}
b_n=\sum_{m=1}^{M}x_{mn},\quad b_n\ge 1,\quad \forall n. \label{eq:C_b_def}
\end{align}
The condition $b_n\ge 1$ enforces valid fixed-$N$ batching, i.e., each declared batch is active.

Batch-level max variables are represented by epigraph inequalities:
\begin{align}
L_n\ge l_m x_{mn},\quad \forall m,n, \label{eq:C_L_epi}\\
I_n\ge i_m x_{mn},\quad \forall m,n. \label{eq:C_I_epi}
\end{align}
Thus, $(L_n,I_n)$ act as batch-wise maxima of speculative length and prefix length.

\subsection{Latency-Coupling Constraints}
The per-user drafting latency is defined as:
\begin{align}
\tau_m^{\mathrm d}=l_mG_m\left(1,F^{\mathrm d}(i_m)\right),\quad \forall m. \label{eq:C_tau_d}
\end{align}
This formulation reflects that the drafting phase is executed locally on each device without batching. Specifically, as the drafting process is inherently serial, the total latency is calculated as the product of the draft length $l_m$ and the per-token drafting time.

Batch draft time uses max-epigraph constraints:
\begin{align}
t_n^{\mathrm d}\ge x_{mn}\!\left(\tau_m^{\mathrm d}+\tau_m^{\mathrm c}\right),\quad \forall m,n. \label{eq:C_td}
\end{align}
so $t_n^{\mathrm d}$ captures when all requests in batch $n$ have completed drafting and communication.

Batch verification latency is
\begin{align}
t_n^{\mathrm v}=G^{\mathrm v}\!\left(b_n,F^{\mathrm v}(L_n,I_n)\right),\quad \forall n. \label{eq:C_tv}
\end{align}
which depends on the batch size $b_n$ and the batch-wise maxima $(L_n,I_n)$ due to padding.
Stage duration must cover verification:
\begin{align}
T_n\ge t_n^{\mathrm v},\quad \forall n. \label{eq:C_T_lb}
\end{align}
Here $T_n$ is the \emph{stage duration} in the pipeline schedule: it can exceed $t_n^{\mathrm v}$ to satisfy global precedence constraints and account for pipeline bubbles.

The total pipeline span $S$ is the steady-state cycle time over which one complete SD round (all $N$ batches) is processed:
\begin{align}
S=\sum_{n=1}^{N}T_n. \label{eq:C_S_def}
\end{align}
Pipeline precedence requires the pipeline span to cover both the draft completion time and the verification latency of batch $n$:
\begin{align}
S \ge t_n^{\mathrm d} + t_n^{\mathrm v},\quad \forall n. \label{eq:C_precedence}
\end{align}

\subsection{Memory Constraints at Edge Side}
Since batching is not employed on the user side, we assume that the memory footprint of the draft model remains well within the device's limits. Consequently, our analysis focuses on the memory constraints at the edge server, which performs batch processing for the target model inference.
With FP16 precision, the model parameters occupy a memory footprint of approximately:
\begin{align}
\Gamma_{\mathrm p}^{\mathrm v}=J_{\mathrm v}\!\left(8(h_1^{\mathrm v})^2+4h_1^{\mathrm v}h_2^{\mathrm v}\right). \label{eq:C_mem_param}
\end{align}
KV-cache component for batch $n$ is:
\begin{align}
\Gamma_{\mathrm{kv},n}^{\mathrm v}=4J_{\mathrm v}h_1^{\mathrm v}b_n I_n,\quad \forall n. \label{eq:C_mem_kv}
\end{align}
The memory feasibility constraint is then:
\begin{align}
\Gamma_{\mathrm p}^{\mathrm v}+\Gamma_{\mathrm{kv},n}^{\mathrm v}\le \Gamma_{\max}^{\mathrm v},\quad \forall n. \label{eq:C_mem_total}
\end{align}

\subsection{Overall Joint Optimization Objective}
We define the throughput $R$ as the expected number of accepted tokens per unit time:
\begin{align}
R \triangleq \frac{U(\mathbf{l})}{S}, \label{eq:R_def}
\end{align}
where $U(\mathbf{l})$ is defined in \eqref{eq:U_def} and $S$ is the total pipeline span in \eqref{eq:C_S_def}.
This formulation is valid under the stationary workload assumption: since the pipeline schedule repeats across successive SD rounds within each optimization period, it suffices to evaluate throughput over a single representative round.

The final design jointly optimizes assignment $\mathbf{x}$, speculative lengths $\mathbf{l}$, batch count $N$, and other auxiliary variables:
\begin{align}
\mathcal{P}:\quad
\max_{\mathbf{x},\mathbf{l},N,\mathrm{aux}} \;
&R \label{eq:global_ratio_obj}\\
\text{s.t.}\;
&\eqref{eq:C_x_binary}\text{--}\eqref{eq:C_mem_total},\nonumber\\
&N\in\{2,\ldots,M\}. \label{eq:global_ratio}
\end{align}

\section{Proposed Joint Solver}

\subsection{Joint Solver Overview}
The joint design problem is challenging because it involves (i) integer decisions (user association and draft length) and (ii) a throughput objective that is fractional in the total latency. We solve it with a two-level joint solver. At the outer level, we scan the number of batches $N\in\{2,\ldots,M\}$. For each fixed $N$, we solve
\begin{align}
\mathcal{P}(N):\quad
\max_{\mathbf{x},\mathbf{l},\mathrm{aux}} \;
&R(N) \label{eq:fixedN_ratio_obj}\\
\text{s.t.}\;
&\eqref{eq:C_x_binary}\text{--}\eqref{eq:C_mem_total}. \nonumber
\end{align}
where $R(N)$ is the throughput defined in \eqref{eq:R_def} under fixed $N$.
To handle the coupling between user-to-batch association and integer draft lengths, we apply alternating optimization:
\begin{align}
\mathbf{x}^{(r+1)} &\leftarrow \arg\min S(N,\mathbf{l}^{(r)}), \label{eq:alt_x}\\
\mathbf{l}^{(r+1)} &\leftarrow \arg\max R(N,\mathbf{x}^{(r+1)}). \label{eq:alt_lm}
\end{align}
The outer loop stops when $\mathbf{l}$ is unchanged and the throughput change is below tolerance.
Finally, we select the best solution over the scanned batch counts. The overall workflow is summarized in Algorithm~\ref{alg:dipsd}. Each subproblem reduces to a mixed-integer linear program (MILP) that we solve using SCIP through its Python interface PySCIPOpt \cite{Achterberg2009SCIP,MaherMiltenbergerPedrosoRehfeldtSchwarzSerrano2016}.

In implementation, we additionally impose practical variable bounds to tighten the MILPs and improve numerical stability. In particular, we use the bound parameters $(L_{\mathrm{ub}},I_{\mathrm{ub}},T_{\mathrm{ub}},S_{\mathrm{ub}})$ as upper bounds for draft lengths, prefix lengths, and auxiliary timing variables.

\subsection{$\mathbf{x}$-Subproblem}
Given fixed $\mathbf{l}$, we solve
\begin{align}
\mathcal{X}(N,\mathbf{l}):\quad
\min_{\mathbf{x},\mathbf{T},\mathrm{aux}} \;
&S \quad \\
\text{s.t. } \;
&\eqref{eq:C_x_binary}\text{--}\eqref{eq:C_x_assign},\eqref{eq:C_b_def}\text{--}\eqref{eq:C_I_epi},\nonumber\\
&\eqref{eq:C_td}\text{--}\eqref{eq:C_mem_total}. \nonumber
\end{align}
This step decides batching assignment and batch-wise timing variables under full feasibility constraints.

\subsection{$\mathbf{l}$-Subproblem: Binary Reformulation + Dinkelbach}
Given fixed $\mathbf{x}$, we solve
\begin{align}
\mathcal{L}(N,\mathbf{x}):\quad
\max_{\mathbf{l}} \;
&\frac{U(\mathbf{l})}{S(\mathbf{l}\mid\mathbf{x})}
\\
\text{s.t.}\;
&\eqref{eq:C_l_integer},\ \eqref{eq:C_L_epi},\ \eqref{eq:C_tau_d}\text{--}\eqref{eq:C_mem_total}. \nonumber
\end{align}
We constrain the draft length to $l_m \in \{1, \ldots, L_{\mathrm{ub}}\}$, where $L_{\mathrm{ub}}$ denotes the upper bound. This constraint is justified by the diminishing returns of the utility function $u_m(l_m)$ as $l_m$ increases. Furthermore, empirical evidence from prior speculative decoding studies \cite{leviathan2023specdec, chen2023specsampling} suggests that the optimal draft length typically falls within the range of 5--15.
To handle integer draft lengths efficiently, we use binary selectors:
\begin{align}
&y_{mk}\in\{0,1\},\quad \sum_{k=1}^{L_{\mathrm{ub}}}y_{mk}=1,\quad
l_m=\sum_{k=1}^{L_{\mathrm{ub}}}k\,y_{mk}, \label{eq:y_def}\\
&z_{nk}\in\{0,1\},\quad \sum_{k=0}^{L_{\mathrm{ub}}}z_{nk}=1,\quad
L_n=\sum_{k=0}^{L_{\mathrm{ub}}}k\,z_{nk}. \label{eq:z_def}
\end{align}
For users in batch $n$, max-consistency is enforced by
\begin{align}
L_n\ge l_m,\quad \forall m\in\mathcal{B}_n(\mathbf{x}),\ \forall n, \label{eq:Lmax_cons}
\end{align}
where $\mathcal{B}_n(\mathbf{x})$ denotes the user set of the $n$-th batch.

Define lookup utility
\begin{align}
u_{mk}=\frac{1-\alpha_m^{k+1}}{1-\alpha_m},\quad
U(\mathbf{y})=\sum_{m=1}^{M}\sum_{k=1}^{L_{\mathrm{ub}}}u_{mk}y_{mk}. \label{eq:util_lookup}
\end{align}
We handle the fractional objective via the classic Dinkelbach method\cite{dinkelbach1967nonlinear}.
At Dinkelbach iteration $t$, solve inner mixed-integer problem:
\begin{align}
\max_{\mathbf{y},\mathbf{z},\mathrm{aux}} \label{eq:dinkel}
\;
&U(\mathbf{y})-q^{(t)}S(\mathbf{y},\mathbf{z})
\\
\text{s.t.}\; &\eqref{eq:C_tau_d}\text{--}\eqref{eq:C_mem_total},\ \eqref{eq:y_def}\text{--}\eqref{eq:Lmax_cons}. \nonumber
\end{align}
Then update
\begin{align}
q^{(t+1)}=\frac{U(\mathbf{y}^{(t)})}{S(\mathbf{y}^{(t)},\mathbf{z}^{(t)})}. \label{eq:dinkel_update}
\end{align}
Stop when
\begin{align}
\left|U(\mathbf{y}^{(t)})-q^{(t)}S(\mathbf{y}^{(t)},\mathbf{z}^{(t)})\right|\le \epsilon, \label{eq:dinkel_stop}
\end{align}
or an equivalent throughput-change threshold is met.
Each inner problem is a MILP. We solve these MILPs using SCIP/PySCIPOpt \cite{Achterberg2009SCIP,MaherMiltenbergerPedrosoRehfeldtSchwarzSerrano2016}.

\subsection{Batch-Count Selection and Final Deployment Decision}
For each fixed $N$, let the converged throughput be $R^\star(N)$. Then we select
\begin{align}
N^\star=\arg\max_{N\in\{2,\ldots,M\}}R^\star(N). \label{eq:batch_select}
\end{align}



\begin{algorithm}[t]
\caption{DiP-SD: Joint Optimization Solver}
\label{alg:dipsd}
\KwRequire{$M$, parameter set, $L_{\mathrm{ub}}$, tolerances, solver limits}
\KwEnsure{$N^\star,\mathbf{x}^\star,\mathbf{l}^\star,R^\star$}
\For{$N\leftarrow 2$ \KwTo $M$}{
    Initialize $\mathbf{l}^{(0)}$ and $r\leftarrow 0$\;
    \Repeat{convergence}{
        Solve $\mathbf{x}$-subproblem with fixed $\mathbf{l}^{(r)}$ to get $\mathbf{x}^{(r+1)}$\;
        Solve $\mathbf{l}$-subproblem using \eqref{eq:dinkel}--\eqref{eq:dinkel_stop} to get $\mathbf{l}^{(r+1)}$\;
        $r\leftarrow r+1$\;
    }
    Compute converged throughput $R^\star(N)$\;
}
Select $N^\star=\arg\max_{N\in\{2,\ldots,M\}}R^\star(N)$\;
\Return $(N^\star,\mathbf{x}^\star,\mathbf{l}^\star,R^\star)$\;
\end{algorithm}

\subsection{Complexity Discussion}
As shown in Algorithm~\ref{alg:dipsd}, DiP-SD scans the batch count and alternates MILP-based $\mathbf{x}$ and $\mathbf{l}$ updates for each fixed $N$ to maximize throughput. Let $I_{\mathrm{out}}$ be the number of alternating iterations and $I_{\mathrm{din}}$ be Dinkelbach iterations. Per fixed $N$, complexity is approximately
\[
\mathcal{O}\!\left(I_{\mathrm{out}}\big(\mathcal{C}_{\mathbf{x}}(M,N)+I_{\mathrm{din}}\mathcal{C}_{\mathbf{l}}(M,N,L_{\mathrm{ub}})\big)\right),
\]
where $\mathcal{C}_{\mathbf{x}}$ and $\mathcal{C}_{\mathbf{l}}$ denote solver-dependent inner complexities.
The outer batch-count scan multiplies this by the number of tested batch counts.

\section{Numerical Results}

\subsection{System Setup}
We evaluate DiP-SD using the Qwen3 model series\cite{yang2025qwen3}.
Specifically, Qwen3-1.7B serves as the draft model on an NVIDIA GeForce RTX~3090,
and Qwen3-32B serves as the target model on an NVIDIA A100-80GB edge server.
Their transformer dimensions, $(J_{\mathrm d},h_1^{\mathrm d},h_2^{\mathrm d})$
and $(J_{\mathrm v},h_1^{\mathrm v},h_2^{\mathrm v})$ respectively, along with all other default
parameters, are listed in Table~\ref{tab:param}.
Latency parameters are obtained by profiling each model on its
respective hardware and fitting the measurements to the affine
latency model in~\eqref{eq:Gm}--\eqref{eq:Gv}.
The default acceptance rate is estimated from inference statistics over 100 sampled prompts.
All remaining experiments are conducted via numerical simulation
under the default parameters in Table~\ref{tab:param}.
All latency parameters are in milliseconds~(ms) and memory parameters are in bytes~(B).

We compare five methods:
\begin{itemize}
    \item \textbf{DiP-SD (proposed):} joint solver that scans batch counts and alternates $\mathbf{x}/\mathbf{l}$ updates.
    \item \textbf{AD:} edge-only large model autoregressive decoding without speculative drafting. No batching scheme is applied.
    \item \textbf{AD w/ greedy batching:} AD with greedy memory-feasible batching: sort users by $i_m$ in descending order, then pack them into batches sequentially until the memory bound would be violated.
    \item \textbf{DiP-SD w/o batching:} enforce $N=M$ and one user per batch; optimize only $\mathbf{l}$.
    \item \textbf{DiP-SD w/ fixed $l_m=7$:} fix $l_m$ and optimize only $\mathbf{x}$.
\end{itemize}

The primary metric is throughput, which is expected accepted tokens per unit time in our context. 
Since $S$ is in ms, $R=U(\mathbf{l})/S$ has unit token/ms, and we report throughput with $1000\times R$ token/s.

\begin{table}[t]
\caption{Default system parameters.}
\label{tab:param}
\centering
\begin{tabular}{lc}
\toprule
Parameter & Value \\
\midrule
User number $M$ & $6$ \\
Scan range of $N$ & $2$ to $M$ \\
Acceptance rate $\alpha_m$ & $0.78$ \\
Prefix length $i_m$ & $512$ \\
Communication latency $\tau_m^{\mathrm c}$ & $3.0$ \\
User-side latency parameters $(c_m,\beta_m)$ & $(4.0305\times10^{-11},\,33.8151)$ \\
Initial draft length $l_m$ & $7$ \\
Draft model $(J_{\mathrm d},h_1^{\mathrm d},h_2^{\mathrm d})$ & $(28,\,2048,\,6144)$ \\
Verify model $(J_{\mathrm v},h_1^{\mathrm v},h_2^{\mathrm v})$ & $(64,\,5120,\,25600)$ \\
Verify timing $(c^{\mathrm v},\beta^{\mathrm v})$ & $(1.2077\times10^{-11},\,95.1074)$ \\
Memory cap $\Gamma_{\max}^{\mathrm v}$ & $8.0\times10^{10}$ \\
Bounds $(L_{\mathrm{ub}},I_{\mathrm{ub}},T_{\mathrm{ub}},S_{\mathrm{ub}})$ & $(20,\,1024,\,10^6,\,3\times10^6)$ \\
\bottomrule
\end{tabular}
\end{table}

\subsection{Results}

\subsubsection{Homogeneous Setting}
We first consider a homogeneous-user setting, where all users share the same per-user parameters at each test point. We conduct two groups of experiments: 1) throughput versus user number $M$ (number of concurrent users), with $M\in\{2,4,\ldots,14\}$; and 2) throughput versus token acceptance rate $\alpha_m$, with $M=6$, $\alpha_m\in\{0.70,0.75,0.80,0.85,0.90,0.95\}$. 
Other parameters are set as the default values in Table~\ref{tab:param}.

\begin{figure}[t]
\centering
\subfloat[Throughput vs. user number $M$]{
\includegraphics[width=0.7\columnwidth]{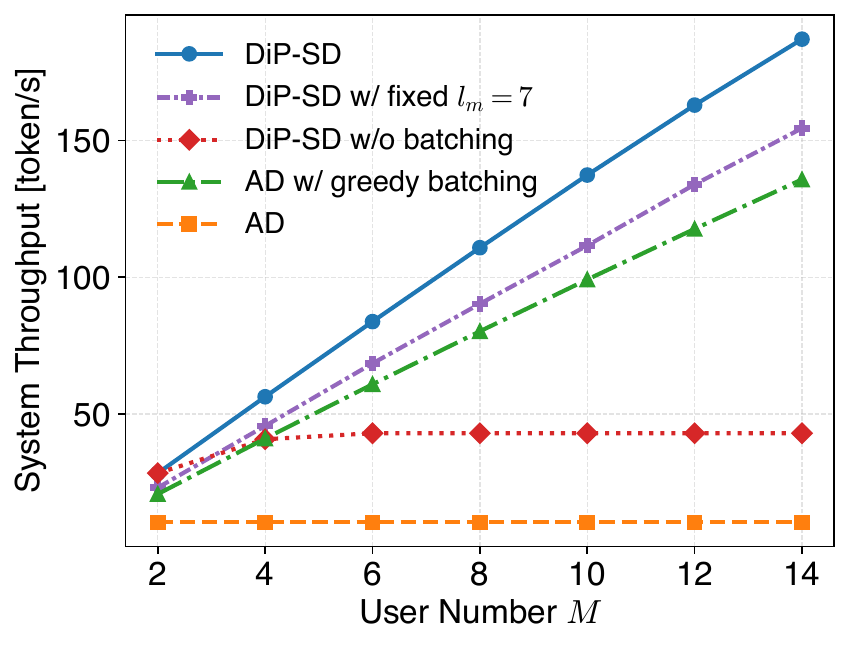}
\label{fig:ratio_m_homo}
}
\vspace{4pt}
\subfloat[Throughput vs. acceptance rate $\alpha_m$ ($M=6$)]{
\includegraphics[width=0.7\columnwidth]{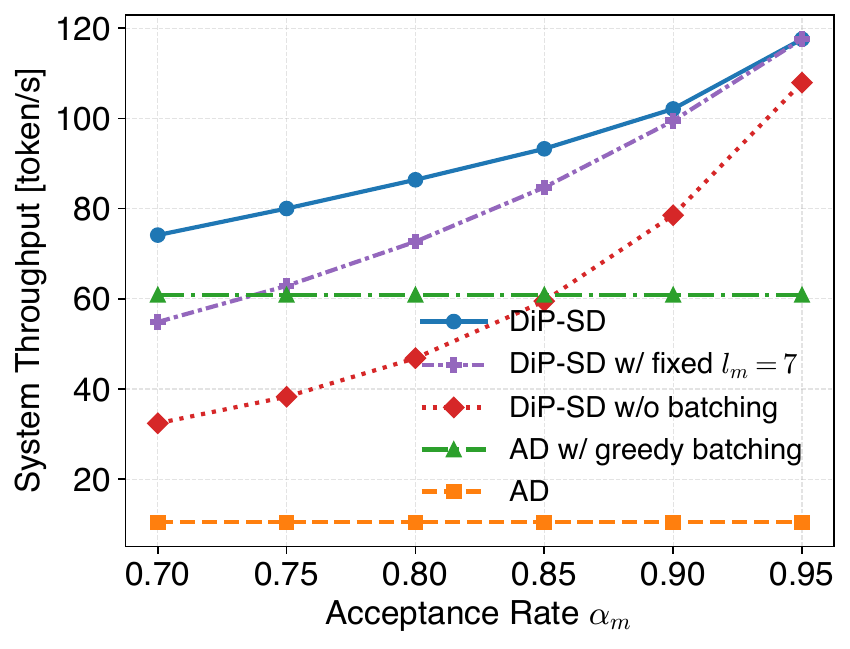}
\label{fig:ratio_alpha_homo}
}
\caption{Homogeneous-setting throughput comparison (token/s).}
\label{fig:homo_results}
\end{figure}

Fig.~\ref{fig:homo_results}(a) shows that the proposed DiP-SD consistently
achieves the highest throughput among all methods.
The basic AD scheme, lacking batching, has throughput independent of user count.
DiP-SD without batching exhibits throughput that first increases with $M$ and then plateaus.
In contrast, batching-enabled methods --- DiP-SD, DiP-SD with fixed $l_m=7$,
and AD with greedy batching --- all exhibit approximately linear throughput growth with increasing $M$.
This is because the throughput numerator scales proportionally with $M$, while newly added
users are absorbed into existing batches with negligible increase in total pipeline span $S$.
The performance gap between DiP-SD and AD with greedy batching widens as $M$ increases;
at peak, DiP-SD achieves $1.38\times$ over AD with greedy batching and $17.89\times$ over AD.
Meanwhile, DiP-SD with fixed $l_m=7$ underperforms DiP-SD,
highlighting the importance of draft length selection.
Furthermore, DiP-SD without batching is substantially outperformed by the three
batching-enabled methods when $M \geq 6$, demonstrating the critical role of batching
in supporting concurrent multi-user inference.

As illustrated in Fig.~\ref{fig:homo_results}(b), the throughput of all SD-based benchmarks exhibits a monotonic increase with the acceptance rate $\alpha_m$. This is because a higher $\alpha_m$ allows more drafted tokens accepted per round. In contrast, the throughput of AD baselines remains constant, as they do not leverage speculative execution.
The proposed DiP-SD scheme maintains the highest throughput at all testpoints. 
Quantitatively, as $\alpha_m$ increases from $0.70$ to $0.95$ (with $M=6$), the throughput of DiP-SD grows from $74.13$ to $117.56$ token/s, representing a $1.59\times$ improvement. 
At its peak, the proposed DiP-SD achieves a substantial speedup of $11.25\times$ over the vanilla AD and $1.93\times$ over the AD baseline with greedy batching, demonstrating the superiority of our method in high-acceptance regimes.
However, for model pairs with a large size gap, the acceptance rate is usually moderate.
In our setting, the default $\alpha_m$ is $0.78$, at which DiP-SD achieves
around $1.38\times$ throughput over AD with greedy batching.
It is also noted that DiP-SD with fixed $l_m=7$ and DiP-SD without batching do not always outperform
AD with greedy batching: their \emph{performance crossover} points occur near
$\alpha_m = 0.75$ and $\alpha_m = 0.85$, respectively, below which the overhead
of failed drafting offsets the gains of speculation.
This further confirms that adaptive draft length and batching each make an essential
contribution to throughput improvement.

\subsubsection{Heterogeneous Setting}
We next evaluate three single-factor heterogeneous cases with fixted user number $M=6$. Case 1 heterogenizes prefix length $i_m$ with values $\{200, 320, 440, 560, 680, 800\}$, while Case 2 considers non-uniform acceptance rates $\alpha_m$ ranging from $\{0.80, 0.82, 0.84, 0.86, 0.88, 0.90\}$. In Case 3, we heterogenize the draft-side latency coefficients $(c_m,\beta_m)$ by applying the scale set $\{0.2, 0.5, 1.0, 3.0, 5.0, 7.0\}$ to the default values. 

\begin{figure}[t]
\centering
\includegraphics[width=0.84\columnwidth]{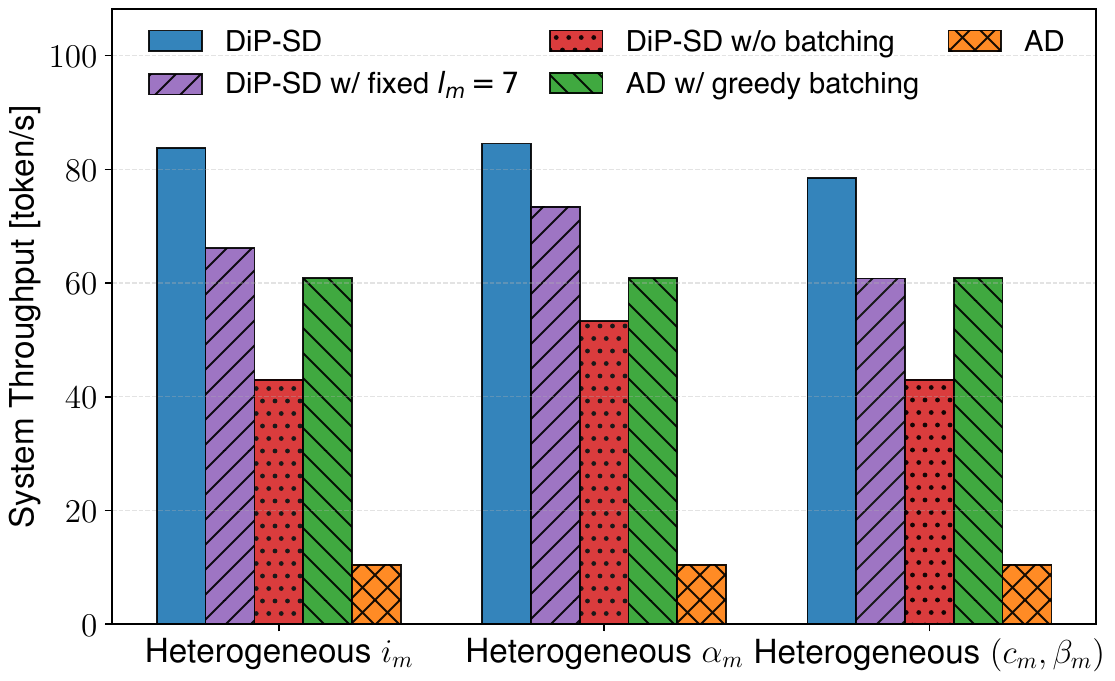}
\caption{Heterogeneous-setting throughput comparison over three single-factor cases (token/s).}
\label{fig:hetero_results}
\end{figure}

Fig.~\ref{fig:hetero_results} shows that DiP-SD remains the best-performing method
across all three heterogeneous cases, achieving up to $8.09\times$ throughput over AD
and $1.39\times$ over AD with greedy batching.
The proposed throughput ranges from $78.49$ to $84.53$~token/s across the three scenarios,
with the $\alpha_m$-heterogeneous case yielding the largest advantage.
This gain is primarily attributed to the inclusion of several users whose acceptance rates
exceed the default value, which directly raises the expected number of accepted tokens per
verification round.
From an architectural perspective, variations in $\alpha_m$ fundamentally affect the utility
of speculation, whereas heterogeneity in prefix length or latency coefficients primarily
impacts the latency components of the system.
It is also worth noting that DiP-SD-based methods suffer a pronounced throughput degradation
under heterogeneous $(c_m,\beta_m)$: the imbalance across user-side latencies disrupts the
pipeline, enlarging the total span and reducing pipeline efficiency.

\section{Conclusion}
This paper presented DiP-SD, a joint optimization framework for distributed
pipelined speculative decoding. We maximize throughput, defined as the
expected number of accepted tokens per unit time, under latency, memory,
and communication constraints. DiP-SD jointly optimizes user-to-batch
association and integer draft lengths via an outer scan over the batch
count and an alternating update between the two subproblems. Numerical
results show that DiP-SD consistently achieves the highest throughput
across all evaluated scenarios, with substantial gains over baselines.

DiP-SD targets edge-first deployments, where multiple devices host
lightweight drafting and a nearby edge server performs shared verification.
This makes the framework suitable for edge assistants, vehicle--road--edge
collaborative inference, and other low-latency interactive LLM applications.
Beyond edge serving, the framework also applies to general multi-accelerator
deployments with distributed drafting and shared verification, covering both
single-node and multi-node GPU configurations. Future work includes
real-world deployment and online adaptation to nonstationary workloads.



\section{Acknowledgement}
This work is supported in part by the State Key Laboratory of Internet of Things for Smart City (University of Macau) Open Research Project under Grant SKL-IoTSC(UM)/ORP04/2026, and in part by the Project of Tsinghua University-Toyota Joint Research Center for AI Technology of Automated Vehicle under Grant TTAD-2024-08-2.

\bibliographystyle{ieeetr}
\bibliography{refs}

@InProceedings{leviathan2023specdec,
  title     = {Fast Inference from Transformers via Speculative Decoding},
  author    = {Leviathan, Yaniv and Kalman, Matan and Matias, Yossi},
  booktitle = {Proceedings of the 40th International Conference on Machine Learning (ICML)},
  year      = {2023}
}

@article{chen2023specsampling,
  title   = {Accelerating Large Language Model Decoding with Speculative Sampling},
  author  = {Chen, Charlie and Borgeaud, Sebastian and Irving, Geoffrey and Lespiau, Jean-Baptiste and Sifre, Laurent and Jumper, John},
  journal = {arXiv preprint arXiv:2302.01318},
  year    = {2023}
}

@article{yang2025qwen3,
  author  = {Yang, A. and Li, A. and Yang, B. and others},
  title   = {Qwen3 Technical Report},
  journal = {arXiv preprint arXiv:2505.09388},
  year    = {2025}
}

@inproceedings{xia2023specdec,
  title     = {Speculative Decoding: Exploiting Speculative Execution for Accelerating Seq2seq Generation},
  author    = {Xia, Heming and Ge, Tao and Wang, Peiyi and Chen, Si-Qing and Wei, Furu and Sui, Zhifang},
  booktitle = {Findings of the Association for Computational Linguistics: EMNLP 2023},
  year      = {2023}
}

@inproceedings{zhou2024distillspec,
  title     = {DistillSpec: Improving Speculative Decoding via Knowledge Distillation},
  author    = {Zhou, Yongchao and Lyu, Kaifeng and Rawat, Ankit Singh and Menon, Aditya Krishna and Rostamizadeh, Afshin and Kumar, Sanjiv and Kagy, Jean-Francois and Agarwal, Rishabh},
  booktitle = {International Conference on Learning Representations (ICLR)},
  year      = {2024}
}

@InProceedings{liu2024osd,
  title     = {Online Speculative Decoding},
  author    = {Liu, Xiaoxuan and Hu, Lanxiang and Bailis, Peter and Cheung, Alvin and Deng, Zhijie and Stoica, Ion and Zhang, Hao},
  booktitle = {Proceedings of the 41st International Conference on Machine Learning (ICML)},
  year      = {2024}
}

@InProceedings{li2024eagle,
  title     = {{EAGLE}: Speculative Sampling Requires Rethinking Feature Uncertainty},
  author    = {Li, Yuhui and Wei, Fangyun and Zhang, Chao and Zhang, Hongyang},
  booktitle = {Proceedings of the 41st International Conference on Machine Learning (ICML)},
  year      = {2024}
}

@inproceedings{li2024eagle2,
  author    = {Li, Yuhui and Wei, Fangyun and Zhang, Chao and Zhang, Hongyang},
  title     = {{EAGLE-2}: Faster Inference of Language Models with Dynamic Draft Trees},
  booktitle = {Proceedings of the 2024 Conference on Empirical Methods in Natural Language Processing (EMNLP)},
  year      = {2024}
}

@InProceedings{cai2024medusa,
  title     = {Medusa: Simple {LLM} Inference Acceleration Framework with Multiple Decoding Heads},
  author    = {Cai, Tianle and Li, Yuhong and Geng, Zhengyang and Peng, Hongwu and Lee, Jason D. and Chen, Deming and Dao, Tri},
  booktitle = {Proceedings of the 41st International Conference on Machine Learning (ICML)},
  year      = {2024}
}

@inproceedings{zhang2024draftverify,
  title     = {Draft {\&} Verify: Lossless Large Language Model Acceleration via Self-Speculative Decoding},
  author    = {Zhang, Jun and Wang, Jue and Li, Huan and Shou, Lidan and Chen, Ke and Chen, Gang and Mehrotra, Sharad},
  booktitle = {Proceedings of the 62nd Annual Meeting of the Association for Computational Linguistics (ACL)},
  year      = {2024}
}

@inproceedings{li2025sled,
  title     = {SLED: A Speculative LLM Decoding Framework for Efficient Edge Serving},
  author    = {Li, Xiangchen and Spatharakis, Dimitris and Ghafouri, Saeid and Fan, Jiakun and Vandierendonck, Hans and John, Deepu and Ji, Bo and Nikolopoulos, Dimitrios S.},
  booktitle = {Proceedings of the Tenth ACM/IEEE Symposium on Edge Computing (SEC)},
  year      = {2025}
}

@inproceedings{chen2025spin,
  title     = {SPIN: Accelerating Large Language Model Inference with Heterogeneous Speculative Models},
  author    = {Chen, Fahao and Li, Peng and Luan, Tom H. and Su, Zhou and Deng, Jing},
  booktitle = {Proceedings of the 44th IEEE International Conference on Computer Communications (INFOCOM)},
  year      = {2025}
}

@inproceedings{ning2025dssd,
  title     = {{DSSD}: Efficient Edge-Device Deployment and Collaborative Inference via Distributed Split Speculative Decoding},
  author    = {Ning, Jiahong and Zheng, Ce and Yang, Tingting},
  booktitle = {ICML 2025 Workshop on Machine Learning for Wireless Communication and Networks},
  year      = {2025}
}

@article{venkatesha2025earlyexit,
  title   = {Fast and Cost-effective Speculative Edge-Cloud Decoding with Early Exits},
  author  = {Venkatesha, Yeshwanth and Kundu, Souvik and Panda, Priyadarshini},
  journal = {arXiv preprint arXiv:2505.21594},
  year    = {2025}
}

@article{zheng2025tkslt,
  title   = {Communication-Efficient Collaborative LLM Inference via Distributed Speculative Decoding},
  author  = {Zheng, Ce and Yang, Tingting},
  journal = {arXiv preprint arXiv:2509.04576},
  year    = {2025}
}

@article{bhattacharjee2025conformal,
  title   = {Conformal Sparsification for Bandwidth-Efficient Edge-Cloud Speculative Decoding},
  author  = {Bhattacharjee, Payel and Tian, Fengwei and Zhong, Meiyu and Zhang, Guangyi and Simeone, Osvaldo and Tandon, Ravi},
  journal = {arXiv preprint arXiv:2510.09942},
  year    = {2025}
}

@article{liu2025flowspec,
  title   = {FlowSpec: Continuous Pipelined Speculative Decoding for Efficient Distributed LLM Inference},
  author  = {Liu, Xing and Luo, Lizhuo and Tang, Ming and Huang, Chao and Chen, Xu},
  journal = {arXiv preprint arXiv:2507.02620},
  year    = {2026}
}

@article{yu2025dsd,
  title   = {DSD: A Distributed Speculative Decoding Solution for Edge-Cloud Agile Large Model Serving},
  author  = {Yu, Fengze and Li, Leshu and McDanel, Brad and Zhang, Sai Qian},
  journal = {arXiv preprint arXiv:2511.21669},
  year    = {2025}
}

@article{li2026flexspec,
  title   = {FlexSpec: Frozen Drafts Meet Evolving Targets in Edge-Cloud Collaborative LLM Speculative Decoding},
  author  = {Li, Yuchen and Kong, Rui and Lyu, Zhonghao and Li, Qiyang and Chen, Xinran and Cai, Hengyi and Yan, Lingyong and Wang, Shuaiqiang and Zhao, Jiashu and Zhu, Guangxu and Kong, Linghe and Chen, Guihai and Xiong, Haoyi and Yin, Dawei},
  journal = {arXiv preprint arXiv:2601.00644},
  year    = {2026}
}

@article{zhu2025efficient,
  title   = {Efficient LLM Inference over Heterogeneous Edge Networks with Speculative Decoding},
  author  = {Zhu, Bingjie and Chen, Zhixiong and Zhao, Liqiang and Shin, Hyundong and Nallanathan, Arumugam},
  journal = {arXiv preprint arXiv:2510.11331},
  year    = {2025}
}

@article{Achterberg2009SCIP,
  title     = {{SCIP}: Solving Constraint Integer Programs},
  author    = {Achterberg, Tobias},
  journal   = {Mathematical Programming Computation},
  volume    = {1},
  number    = {1},
  pages     = {1--41},
  year      = {2009},
  doi       = {10.1007/s12532-008-0001-1},
  publisher = {Springer}
}

@inproceedings{MaherMiltenbergerPedrosoRehfeldtSchwarzSerrano2016,
  title     = {PySCIPOpt: Mathematical Programming in {Python} with the {SCIP} Optimization Suite},
  author    = {Maher, Stephen and Miltenberger, Markus and Pedroso, Jo{\~a}o Pedro and Rehfeldt, Daniel and Schwarz, Robert and Serrano, Felipe},
  booktitle = {Mathematical Software -- ICMS 2016},
  year      = {2016}
}

@inproceedings{liu2025pearl,
  title     = {PEARL: Parallel Speculative Decoding with Adaptive Draft Length},
  author    = {Liu, Tianyu and Li, Yun and Lv, Qitan and Liu, Kai and Zhu, Jianchen and Hu, Winston and Sun, Xiao},
  booktitle = {International Conference on Learning Representations (ICLR)},
  year      = {2025}
}

@article{liu2024turbospec_goodput,
  title   = {Optimizing Speculative Decoding for Serving Large Language Models Using Goodput},
  author  = {Liu, Xiaoxuan and Daniel, Cade and Hu, Langxiang and Kwon, Woosuk and Li, Zhuohan and Mo, Xiangxi and Cheung, Alvin and Deng, Zhijie and Stoica, Ion and Zhang, Hao},
  journal = {arXiv preprint arXiv:2406.14066},
  year    = {2024}
}

@inproceedings{xu2023smdp,
  title     = {SMDP-Based Dynamic Batching for Efficient Inference on GPU-Based Platforms},
  author    = {Xu, Yaodan and Sun, Jingzhou and Zhou, Sheng and Niu, Zhisheng},
  booktitle = {Proceedings of the IEEE International Conference on Communications (ICC)},
  year      = {2023}
}

@article{10829781,
  author  = {Xu, Yaodan and Zhou, Sheng and Niu, Zhisheng},
  journal = {IEEE Transactions on Parallel and Distributed Systems},
  title   = {SMDP-Based Dynamic Batching for Improving Responsiveness and Energy Efficiency of Batch Services},
  year    = {2025},
  volume  = {},
  number  = {},
  pages   = {1--16},
  doi     = {10.1109/TPDS.2025.3526283}
}

@article{dinkelbach1967nonlinear,
  title   = {On Nonlinear Fractional Programming},
  author  = {Dinkelbach, Werner},
  journal = {Management Science},
  volume  = {13},
  number  = {7},
  pages   = {492--498},
  year    = {1967},
  doi     = {10.1287/mnsc.13.7.492}
}

\end{document}